\documentclass[plb,showpacs,preprintnumbers,
a4paper, nofootinbib ,twocolumn,
]{revtex4}
\usepackage{graphicx}

\newcommand{\beq}{\begin{equation}}
\newcommand{\eeq}{\end{equation}}
\newcommand{\beqs}{\begin{eqnarray}}
\newcommand{\eeqs}{\end{eqnarray}}

\begin{document}

\preprint{FERMILAB-PUB-08-001-T}

\title{Mini Little Higgs and Dark Matter}
\author{Yang Bai}
\affiliation{Theoretical Physics Department, Fermilab, Batavia, IL 60510, USA}

\begin{abstract}

We construct a little Higgs model with the most minimal extension of the standard model gauge group by an extra $U(1)$ gauge symmetry. For specific charge assignments of scalars, an approximate $U(3)$ global symmetry appears in the cutoff-squared scalar mass terms generated from gauge bosons at one-loop level. Hence, the Higgs boson, identified as a pseudo-Goldstone boson of the broken global symmetry, has its mass radiatively protected up to scales of 5-10~TeV. In this model, a $Z_2$ symmetry, ensuring the two $U(1)$ gauge groups to be identical, also makes the extra massive neutral gauge boson stable and a viable dark matter candidate with a promising prospect of direct detection. 

\end{abstract}


\maketitle

In the standard model, the electroweak symmetry breaking is described by a Higgs field, whose mass is not stable against radiative corrections, and hence the electroweak scale is not stable either. On the other hand, the electroweak precision measurements require the cutoff scale of the new physics beyond the standard model to be greater than 5-10 TeV. Therefore a delicate fine-tuning is required to explain the observed weak scale. This is known as the little hierarchy problem. One solution to this problem is to identify the Higgs doublet as a pseudo-Goldstone boson as described in little Higgs models~\cite{LittleHiggs}.

However, most of little Higgs models are based on complicated non-linear sigma models, and require a large amount of new particles around the TeV scale. In this letter, we propose a novel little Higgs model, which is as simple as possible and only contains a minimal set of new particles beyond the standard model. To cancel cutoff-squared contributions to the Higgs boson mass from the standard model gauge bosons and without invoking the supersymmetry, we can enlarge the electroweak gauge group of the standard model. The most minimal extension is to introduce an extra $U(1)$ gauge group, and hence we choose the gauge group to be $SU(2)_w\times U(1)_1\times U(1)_2$. So unlike the traditional little Higgs models, there is no new charged gauge boson in our model, and one may wonder what cancels the quadratic divergent part from the $W$ gauge boson. Below, we will show that the cancellation happens among contributions from $W$, $Z$ and the extra neutral gauge boson called $B^\prime$, depending on the $U(1)$ charges of scalars.

We illustrate our basic idea by first looking at the bosonic sector of our model. To make our model more minimal, we assume a $Z_2$ interchanging symmetry between these two $U(1)$ gauge groups by equalizing their gauge couplings. This $Z_2$ symmetry can be identified as the $T$-parity as in~\cite{Cheng:2003ju} to play the same roles as R-parity in the supersymmetric theories and KK-parity in universal extra-dimensions~\cite{Appelquist:2000nn} to provide a dark matter candidate. Under the gauge symmetries, we choose a doublet, $H$, transforming as $(2, \frac{1}{2}, \frac{1}{2})$ and an $SU(2)_w$ singlet, $S$, charged as $(1, \frac{5}{3}, -\frac{5}{3})$, with the charge assignments assumed to come from some unified theories.

From the kinetic terms of scalars,
\beqs
|D_\mu H|^2&=&|(\partial_\mu + i g\,t^a\,W_\mu^a + i\frac{g^\prime}{2\sqrt{2}}(B_{1\mu} +B_{2\mu })) H|^2\,, \nonumber  \\ 
|D_\mu S|^2&=&|(\partial_\mu + i\frac{5g^\prime}{3\sqrt{2}}(B_{1\mu} - B_{2\mu})) S|^2\,, \label{Eq:kinetic}
\eeqs
the cutoff-squared contributions to scalar masses from gauge bosons at one-loop level are 
\beqs
V_g&=&\frac{3\,\Lambda^2}{64 \pi^2}\left[(3 g^2\,+\,g^{\prime 2}) H H^\dagger + \frac{100}{9} g^{\prime 2} S S^\dagger\right]\,+\,\cdots\,, \nonumber \\
&\approx& \frac{25 g^{\prime 2}\Lambda^2}{48 \pi^2} \left[H H^\dagger + S S^\dagger\right]\,+\,\cdots\,. \label{Eq:U3}
\eeqs
Here $g$ is the gauge coupling of $SU(2)_w$ with $t^a$ as its three generators; the gauge couplings of $U(1)$'s are chosen to be identical and are $\sqrt{2}g^\prime$, with $g^\prime$ as the gauge coupling of the $U(1)_Y$ hypercharge symmetry; the experimental value of the weak mixing angle $s^2_w\equiv\sin^2{\theta_w}=g^{\prime 2}/(g^2+g^{\prime 2}) \approx 0.23$ is used in the second line of the above equation, and the coefficients in front of $HH^\dagger$ and $SS^\dagger$ are approximately equal up to a few percent. As can be seen, the quadratically divergent part of the scalar potential satisfies an accidental $U(3)$ global symmetry~\cite{TwinHiggs}. So if the Higgs doublet is identified as a Goldstone boson in $U(3)/U(2)$, the cutoff-squared contributions to the Higgs mass-squared from gauge bosons are then cancelled because of the approximate $U(3)$ global symmetry.

Now we switch to nonlinear descriptions of our model. We write these two scalars together as a triplet of the global symmetry $U(3)$ as $\phi = (H, S)^T$. With the generators of the two $U(1)$'s defined as $T^{4,5}=\frac{1}{n}\,\mbox{diag}(\frac{1}{2}, \frac{1}{2}, \pm\frac{5}{3})$ and the generators of $SU(2)$ as $T^a=\mbox{diag}(t^a, 0)$, the field $\phi$ has the covariant derivative: $D_\mu \phi=(\partial_\mu + i g\,T^a\,W_\mu^a + i g^\prime \frac{n}{\sqrt{2}}\,(T^4\,B_{1\mu} +T^5\,B_{2\mu }))\phi$. Here $n=\sqrt{59}/3$ is the normalization factor to have ${\mbox{tr}}(T^i T^i)=\frac{1}{2}$. Up to global symmetry transformation, $\phi$ is assumed to develop a vacuum expectation value (VEV) as $\langle \phi \rangle = (0, 0, f)^T$ and breaks the global symmetry $U(3)$  to $U(2)$ and the gauge symmetry $SU(2)_w\times U(1)_1\times U(1)_2$ to $SU(2)_w\times U(1)_Y$. Here $f$ is taken to be higher than the electroweak scale $v$, and defines the cutoff of the effective field theory as $\Lambda \approx 4 \pi f$. Among the 5 Goldstone bosons in the effective field theory, one is eaten by the massive neutral gauge boson $B^\prime\equiv (B_1-B_2)/\sqrt{2}$, while the other 4 become pseudo-Goldstone bosons and identified as the Higgs doublet $h$. We parametrize the triplet in terms of $h$ as
\beqs
\phi^T =f (i\,\sin{\frac{ h}{f}},\,\cos{\frac{h}{f}})= (ih,\, f-\frac{h\,h^\dagger}{2f}) +\cdots\,.
\eeqs
From the kinetic terms of scalars, the field dependent masses of gauge bosons are derived as
\beqs
M^2_W(h) &=& c_w^2\,M^2_Z(h) = \frac{1}{2}g^2f^2\sin^2{\frac{h}{f}} \,,\nonumber \\
M^2_{B^\prime}(h) &=&  \frac{50}{9}g^{\prime 2}f^2\cos^2{\frac{ h}{f}}\,,
\eeqs
and used to calculate the one-loop Coleman-Weinberg effective potential~\cite{Coleman:1973jx}, from which we obtain the leading terms of the Higgs potential as $V(h)= m^2_h h h^\dagger + \lambda_h (hh^\dagger)^2+\cdots $~\cite{Appelquist:2005iz}. The Higgs mass contributions from the gauge sector at one-loop level are
\beqs
m^2_{h}|_g=\frac{3g^{\prime 2}\Lambda^2}{32\pi^2}(\frac{27-118s^2_w}{9s^2_w})+\frac{3 M^4_{B^\prime}}{32\pi^2 f^2}(\log{\frac{\Lambda^2}{M^2_{B^\prime}}}+1)\,, \label{Eq:mhg}
\eeqs
with $M_{B^\prime}=5\sqrt{2}g^{\prime}f/3\approx 0.8 f$. The contributions to the Higgs quartic term from the gauge sector are smaller than from the fermion sector, and are neglected here. For $s^2_w$ in the range $(0.22, 0.24)$ \cite{Weinberg} and $\Lambda$ in the range of 5 to 10~TeV, the cutoff-squared contributions to the Higgs mass are cancelled to be smaller than the logarithmically divergent part. This cancellation comes from the $U(1)$ charge assignments of the $SU(2)_w$ singlet $S$. Assuming some grand unified theories can provide us this set of simple discrete choices of $U(1)$ charges, then this accidental cancellation is free from fine-tuning, since we can not continuously change the quantized charges of fields.

In the fermionic sector of our model, all standard model fermions have identical charges under these two $U(1)$'s and have their charges to be one half of their hypercharges. For example, we have $(t_L, b_L)^T$ as $(2, \frac{1}{6}, \frac{1}{6})$, $t_R$ as $(1, \frac{2}{3}, \frac{2}{3})$ and $b_R$ as $(1, -\frac{1}{3}, -\frac{1}{3})$ under the gauge symmetries of our model. Hence our model is free of gauge anomaly. To cancel the cutoff-squared contribution to the Higgs boson mass from the top quark, there are two ways to introduce additional vector-like fermions. The simplest way with only one vector-like fermion manifestly breaks the $Z_2$ symmetry, while the other way with more vector-like fermions keeps the $Z_2$ symmetry  exact. We will consider both cases in turn. 

The most minimal way to cancel  cutoff-squared contribution to the Higgs mass from the top quark is to introduce a colored vector-like quark $\psi_{L,R}$ charged as $(1, \frac{7}{3}, -1)$.  The Yukawa couplings in the top sector are
\beqs
{\cal L}_t&=&y_1(\bar{q}_L\,\tilde{H}\,+\,\bar{\psi}_L\,S)\,t_R\,+\,y_2\,f\,\bar{\psi}_L\,\psi_R\,+\,h.c. \,,
\eeqs
with the first term $U(3)$ invariant and $y_2$ breaking this global symmetry. These two Yukawa couplings in the top sector manifestly break the $Z_2$ symmetry, so $B^\prime$ directly couples to the mass eigenstate of the top quark through the mixing between the top quark and the top partner. Therefore, there is no dark matter candidate in this case. The one-loop contributions to the Higgs boson mass in the top sector are free from cutoff-squared terms, since both couplings $y_1$ and $y_2$ are necessary to generate a potential for the Higgs doublet $h$. We calculate the Higgs doublet mass and quartic coupling as
\beqs
m^2_h|_t&=& -\frac{3}{8\pi^2}y_t^2\,m^2_{t^\prime} (\log{\frac{\Lambda^2}{m^2_{t^\prime}}}+1)  \,,    \label{Eq:mht}    \\
\lambda_h|_t &=& \frac{-m^2_h|_t}{3f^2}+\frac{3y_t^4}{16\pi^2}(\log{\frac{\Lambda^2}{m^2_t}}+\log{\frac{\Lambda^2}{m^2_{t^\prime}}}+\frac{3}{2})\,.  \label{Eq:lamht} 
\eeqs
Here $m_t=y_t\,h$ is the top quark mass after $h$ takes its VEV and $y_t=y_1y_2/\sqrt{y_1^2+y_2^2}$ is the top quark Yukawa coupling. To the leading order in $v/f$, $m_{t^\prime}=\sqrt{y_1^2+y_2^2}f$ is the mass of the top partner.

From Eq.~(\ref{Eq:mht}), the Higgs mass from the top sector is still too large, although only logarithmically divergent terms exist. Additional operators are needed to have the Higgs boson mass below 200~GeV. The simplest way is to include a soft $U(3)$ symmetry breaking operator $\mu^2 H H^\dagger$~\cite{TwinHiggs,Bai:2007tv}, which does not reintroduce quadratically divergent contributions to the Higgs mass. Expanding this operator in terms of $h$, we have
\beqs
m^2_h|_\mu= \mu^2\,,\quad \lambda_h|_\mu= -\frac{\mu^2}{3f^2}\,. \label{Eq:mhmu}
\eeqs

Minimizing the Higgs potential from all contributions in Eqs.~(\ref{Eq:mhg},\ref{Eq:mht},\ref{Eq:lamht},\ref{Eq:mhmu}), the electroweak symmetry is broken with the weak scale $v=246$~GeV and the Higgs boson mass $m_{h_0}=168$~GeV by choosing $f=500$~GeV, $\Lambda=4\pi f\approx 6$~TeV, $y_2=1.56$ and $\mu=351$~GeV. Defining the amount of fine-tuning as a variantion of the weak scale in terms of $\mu$ as $\partial \log{v^2}/\partial \log{\mu^2}$, we have the fine-tuning to be 1 to 8 for this set of numbers. For $\Lambda$ in the range $(2\pi f, 4\pi f)$, the Higgs boson mass can vary from 150 to 170~GeV. 

The corrections to electroweak precision observables first appear at one-loop level, since at tree level only experimentally unmeasured top quark couplings to $W$ and $Z$ bosons are changed. In our model, the strongest constraint on $f$ comes from the $T$ parameter defined in~\cite{Peskin:1991sw}, which is calculated to be positive from the top sector at one-loop level as
\beqs
\alpha T&=& \frac{3y_t^2y_1^2m_t^2}{16\pi^2 y_2^2m^2_{t^\prime}}(\log{\frac{m^2_{t^\prime}}{m^2_t}}-1+\frac{y_1^2}{2y_2^2} )\,,
\eeqs
for $m_t\ll m_{t^\prime}$. The current bounds from PDG~\cite{Yao:2006px} have approximately $\alpha T < 1.2 \times 10^{-3}$ at $95\%$ confidence level for the Higgs mass less than $300$~GeV. For $y_1/y_2 < 3/4$, there is no bound on the symmetry breaking scale $f$ from the $T$ parameter. Hence, $f$ can be as low as $400$~GeV.

For this minimal little Higgs model, only two new fields, $B^\prime$ and $t^\prime$, exist in the effective field theory below 5-10~TeV. They can have masses as light as $300$~GeV and $800$~GeV respectively, and are to be discovered at LHC. In the top sector, the ``collective symmetry breaking mechanism'' in the traditional little Higgs models protects the Higgs mass from receiving cutoff-squared contributions at one-loop level. However, different from previous little Higgs models, collective symmetry breaking mechanism is missing in the gauge boson sector of our model. Fortunately, for the specific charge assignments of scalars and the experimental value of the weak mixing angle, the cutoff-squared contributions to the Higgs mass from $W$, $Z$ and $B^\prime$ are approximately cancelled to be even less than the cutoff-logarithmically dependent contributions. Thus this minimal little Higgs model also stabilizes the weak scale up to 5-10~TeV.

Now we consider a less minimal way to extend the top quark sector, which keeps the $Z_2$ or $T$-parity exact and provides a viable dark matter candidate. We introduce the following colored particles: $q_{1_L}$, $t_R$, $q_{2_L}$, $q^\prime_R$, $\psi_{1_{L,R}}$ and $\psi_{2_{L,R}}$, charged as $(2, \frac{1}{6}, \frac{1}{6})$, $(1, \frac{2}{3}, \frac{2}{3})$, $(2, \frac{1}{6}, \frac{1}{6})$, $(2, \frac{1}{6}, \frac{1}{6})$, $(1, \frac{7}{3}, -1)$ and $(1, -1, \frac{7}{3})$ under $SU(2)_w\times U(1)_1\times U(1)_2$ respectively. Since only three additional vector-like fermions are introduced beyond the standard model, the gauge anomalies are cancelled automatically in this case. To keep the collective symmetry breaking mechanism and to preserve the $T$-parity, the following Yukawa couplings are introduced
\beqs
{\cal L}_t&=&\frac{y_1}{\sqrt{2}}\,(\bar{q}_{1_L}\,\tilde{H} + \bar{\psi}_{1_L}\,S)\,t_R +y_2\,f\,\bar{\psi}_{1_L}\,\psi_{1_R} \nonumber \\
&+& \frac{y_1}{\sqrt{2}}\,(\bar{q}_{2_L}\,\tilde{H} + \bar{\psi}_{2_L}\,S^\dagger)\,t_R +y_2\,f\,\bar{\psi}_{2_L}\,\psi_{2_R}  \nonumber \\
&+& \frac{y_3}{\sqrt{2}}\,f\,(\bar{q}_{1_L}-\bar{q}_{2_L})\,q^\prime_R+h.c. \,. \label{Eq:yukawaT}
\eeqs 
Under the $T$-parity transformation, we have
\beqs
T: &&  q_{1_{L}}\leftrightarrow q_{2_{L}}, \quad \psi_{1_{L,R}}\leftrightarrow \psi_{2_{L,R}}, \quad q^\prime_R\rightarrow -q^\prime_R,
\nonumber \\
&& B_1\leftrightarrow B_2, \quad  S\leftrightarrow S^\dagger\,, 
\eeqs
and all other fields are invariant. The Lagrangian ${\cal L}_t$ and  the covariant kinetic terms of fields are invariant under the $T$-parity~\cite{Hill:2007zv}, and hence all particles in our model are eigenstates of the $T$-parity. 

Diagonalizing the fermion  mass matrix, we have the masses of the $T$-odd particles $t^{\prime}_-$ and $q^\prime$ to be $y_2 f$ and $y_3 f$ respectively. To the leading order in $v/f$, the mass of the $T$-even top partner $t^\prime_+$ is $\sqrt{y_1^2+y_2^2}f$. The top quark is also $T$-even with the top Yukawa coupling as $y_t = y_1y_2/\sqrt{y_2^2+y_1^2}$. The cutoff-squared contribution to the Higgs mass from the top quark is cancelled by the $T$-even top partner. The analyses of the full one-loop Higgs potential and the corrections to the electroweak precision observables are similar to the $T$-parity violating case, and the symmetry breaking scale $f$ can be as low as 400~GeV for a 5~TeV cutoff. 

The $B^\prime$ gauge boson in the $T$-parity invariant model is the lightest $T$-odd particle (LTP) for $y_{2,3}\ge 1$. It can not decay into $T$-even standard model particles, and can serve as a viable dark matter candidate. Different from the littlest Higgs model with $T$-parity, where the LTP is much lighter than the symmetry breaking scale $f$ (around $0.2f$)~\cite{Hubisz:2004ft,Birkedal:2006fz,Chen:2006ie,Martin:2006ss}, the $B^\prime$ in our model is only slightly lighter than $f$ (around $0.8f$). The coupling of two $B^\prime$'s to the Higgs boson is $50g^{\prime 2}v/9$, which is a factor of  $100/9$ larger than the coupling of hypercharge-like gauge bosons to the Higgs boson . The present relic abundance of $B^\prime$ is relating to pair-annihilation rates in the non-relativistic limit by the sum of the quantities, $a(X) = v_r\,\sigma(B^\prime B^\prime \rightarrow X)$, with $v_r$ as the relative speed between $B^\prime$ bosons and $X$ as possible final states. In our model, $B^\prime$ mainly annihilate into pairs of $W$, $Z$, $h_0$ bosons and top quarks. To leading order in $v/f$ and $m_h/M_{B^\prime}$, we have 
\beqs
a(\bar{t}t)&=&\frac{16\pi\alpha^2}{3 \cos^4{\theta_w}}\frac{5^4}{3^4}\frac{y_t^4}{y_2^4}\frac{M^2_{B^\prime}}{(M^2_{B^\prime}+m_{t^{\prime}_-}^2)^2} \,,  \\
a(WW)&=&2a(ZZ)=2a(h_0h_0)=\frac{2\pi\alpha^2}{3 \cos^4{\theta_w}}\frac{5^4}{3^4}\frac{1}{M^2_{B^\prime}} \,. \nonumber
\eeqs
Here $5/3$ is from the coupling among $t^{\prime}_-$, $t^{\prime}_+$ and $B^\prime$, and also indicates a relatively large coupling of two $B^\prime$'s to the Higgs boson; $y_t/y_2$ indicates the mixing between ${t^{\prime}_+}_R$ and $t_R$.

\begin{figure}[htb]
\begin{center}
\includegraphics[width=0.8\linewidth]{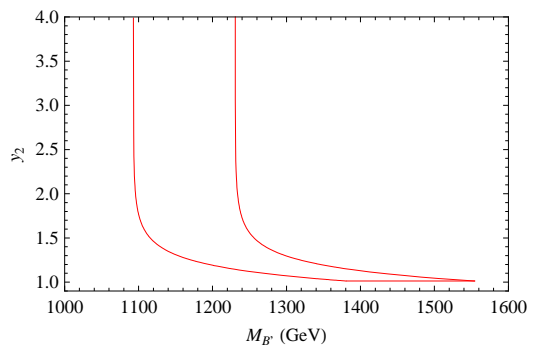}
\caption{The curved box (red) is the allowed region in the $y_2$ and $M_{B^\prime}$ parameter plane to have $B^\prime$ account for all the dark matter with the relic abundance $0.096<\Omega_{B^\prime} h^2 <0.122$ ($2\sigma$) from WMAP. \label{fig:relic}}
\end{center}
\end{figure}

The present dark matter abundance from WMAP collaboration~\cite{Spergel:2006hy}, $0.096<\Omega_{B^\prime} h^2 <0.122$ ($2\sigma$), requires $a_{tot}\approx 0.81\pm 0.09$~pb~\cite{Birkedal:2004xn}, assuming the dark matter candidate $B^\prime$ in our model can make up all the dark matter. In Fig.~\ref{fig:relic}, we plot the allowed region to have $B^\prime$ account for all the dark matter in terms of the parameters $y_2$ and $M_{B^\prime}$ in our model ($y_1$ is not independent and is determined by $y_2$ and $y_t\approx 1$). 
Here $y_2$ is subject to additional constraints $1.013 < y_2 < 4\pi$ to ensure both Yukawa couplings $y_1$ and $y_2$ to be perturbative and below $4\pi$. For $y_2 > 1.5$, $B^\prime$'s mainly annihilate into pairs of bosons, and the relic abundance barely depends on $y_2$. In this region, to account for all the dark matter, the mass of $B^\prime$ is around 1.2~TeV.

\begin{figure}[htb]
\begin{center}
\includegraphics[width=0.9\linewidth]{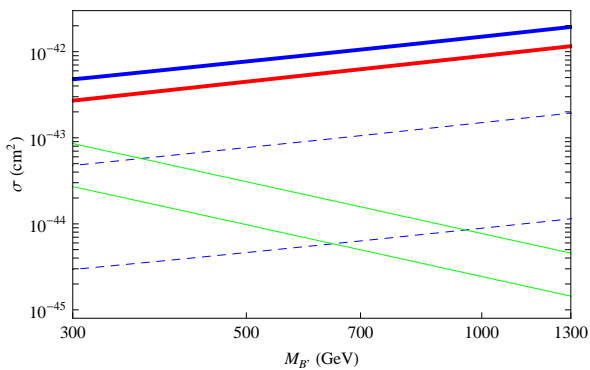}
\caption{Spin-independent $B^\prime$-nucleon elastic scattering cross section. The two thick solid lines are current constraints from CDMS (upper, blue) and XENON (lower, red) experiments. The dashed lines (blue) are projected sensitivities of the CDMS (upper) and the Super-CDMS 25kg (lower) experiments. The thin solid lines (green) are the predictions of this mini little Higgs model for Higgs masses of $120$ (upper) and $160$~GeV (lower) respectively.  \label{fig:DirectDetection}}
\end{center}
\end{figure}

The direct detection of dark matter is to observe the elastic scattering of dark matter particles with nuclei. In our model, the dark matter candidate $B^\prime$ can have elastic scattering with quarks through the t-channel Higgs boson exchange. Therefore, the spin-independent $B^\prime$-nuclei cross section may be measured in our model. (There is also a spin-dependent cross section through a box diagram coupling two $B^\prime$ to two gluons with top quarks and their partners in the loop. For heavy top partners, the spin-dependent cross section is suppressed, and will not be discussed here.) In the non-relativistic limit, the relevent effective operator for the $B^\prime$-quark interaction is $B^\prime_\mu B^{\prime \mu}\bar{q}{q}/2$ with the coefficient as $50g^{\prime 2}m_q/9m_h^2$. Using the matrix elements of quarks in a nucleon state and including the Higgs couplings to gluons mediated by heavy quark loops~\cite{MatrixElement}, we have the spin-independent $B^\prime$-nucleon elastic scattering cross section as
\beqs
\sigma_{SI}&\approx& 1.6\times 10^{-44}\mbox{cm}^2\,(\frac{\mbox{1TeV}}{M_{B^\prime}})^2(\frac{100\mbox{GeV}}{m_h})^4\,.
\eeqs
In Fig.~\ref{fig:DirectDetection}, 
we compare the predicted spin-independent $B^\prime$-nucleon elastic scattering cross section in our model with several experiments. From Fig.~\ref{fig:DirectDetection}, the cross section of direct detection of $B^\prime$ is one to two order of magnitude smaller than the current constraints from CDMS~\cite{CDMS} and XENON~\cite{XENON}, and is accessible by future experiments like the early phase of Super-CDMS~\cite{plotter}. Compared to other hypercharge-like heavy gauge boson dark matter candidates, $\sigma_{SI}$ in our model is two order of magnitude larger~\cite{Ponton:2008bx}. This can be understood from the coupling of $B^\prime$'s to the Higgs boson, which is larger than the coupling of hypercharge-like heavy gauge bosons to the Higgs boson by a factor of $100/9$. This factor is also the crucial factor to cancel the gauge boson cutoff-squared contributions to the Higgs boson mass, and to provide the approximate $U(3)$ global symmetry for the cutoff-squared mass terms in Eq.~(\ref{Eq:U3}).

In conclusion, a very simple little Higgs model has been constructed based on the $SU(2)_w\times U(1)^2$ gauge symmetry. The Higgs boson is identified as a pseudo-Goldstone boson, with its mass radiatively protected up to scales of 5-10 TeV. Depending on vector-like fermion choices in the top sector, a $Z_2$ interchanging symmetry between these two $U(1)$ gauge groups can be broken or unbroken. For the broken case, only a new neutral gauge boson $B^\prime$ and a top partner $t^\prime$ appear in the effective field theory. For the unbroken case, the $B^\prime$ gauge boson is protected by the $Z_2$ symmetry from decaying into standard model fields and can serve as a dark matter candidate. Detailed calculations show that this $B^\prime$ can make up all the dark matter in the universe, and is accessible by the early phase of future dark matter direct detection experiments.

We thank B. Dobrescu, Z. Han, R. Hill, J. Hubisz, K. Kong, J. Lykken and R. Mahbubani for useful discussions.
Fermilab is operated by Fermi Research Alliance, LLC under Contract No.
DE-AC02-07CH11359 with the U.S. Department of Energy.


\end{document}